\documentclass[conference]{IEEEtran}
\IEEEoverridecommandlockouts
\usepackage{cite}
\usepackage{amsmath,amssymb,amsfonts}
\usepackage{algorithmic}
\usepackage{graphicx}
\usepackage{textcomp}
\usepackage{orcidlink}
\usepackage[nolist]{acronym}
\usepackage{graphicx}
\usepackage{xcolor}
\usepackage{xspace}
\usepackage{colortbl}
\usepackage{tikz}
\usepackage{standalone}
\usepackage{booktabs}
\usepackage{footnote}
\makesavenoteenv{tabular}
\usepackage{textpos}
\usepackage{siunitx}
\usepackage{wasysym}
\usepackage{subfig}
\usepackage{tcolorbox}
\usepackage{latexml}

\graphicspath{{fig},{tikzfig}, {fig_results},{arxiv}}

\begin{document}
\def\ie{i.e.}
\def\N{\mathbb N}
\def\R{\mathbb R}
\def\todo#1{\textcolor{red}{#1}}
\def\TM{\texttrademark\ }
\def\1{\mathbf{1}}
\def\0{\mathbf{0}}
\def\I{\mathbb{I}}
\def\mat#1{\ve{#1}}
\def\norm#1{\ensuremath{\left\lVert#1\right\rVert}}
\def\ve#1{\mathlette{\boldmath}{#1}}
\def\pauli#1#2{\ensuremath{\hat{\sigma}_\mathrm{#1}^{(#2)}}}
\def\mathlette#1#2{{\mathchoice{\mbox{#1$\displaystyle #2$}}%
        {\mbox{#1$\textstyle #2$}}%
        {\mbox{#1$\scriptstyle #2$}}%
        {\mbox{#1$\scriptscriptstyle #2$}}}}
\def\x{\ve{x}}
\def\Qxx{\ve{Q}_{xx}}
\def\Qxs{\ve{Q}_{xs}}
\def\Qsx{\ve{Q}_{sx}}
\def\Qss{\ve{Q}_{ss}}
\def\Zx{\ve{Z}_{x}}
\def\Zs{\ve{Z}_{s}} 
\def\tril{\mathrm{tril}}  
\def\triu{\mathrm{triu}}   
\def\Tsim{\ensuremath{T_\mathrm{sim}}}
\def\secref#1{Sec.~#1}
\def\tabref#1{Tab.~#1}
\def\figref#1{Fig.~#1}

\def\colorlabel#1{\tikz{\fill
         (0,0) rectangle (5pt,1pt)}}
     \def\elqq{``}
     \def\erqq{''\xspace}

\DeclareRobustCommand\linelabelblue{\tikz{\node at (0.1,0.025)[]{}; \draw[very thick, blue!60!black!100] (0,0)--(0.3,0) }}
\DeclareRobustCommand\linelabelgreen{\tikz{\node at (0.1,0.025)[]{}; \draw[very thick, green!80!black!100] (0,0)--(0.3,0) }}
\DeclareRobustCommand\linelabelred{\tikz{\node at (0.1,0.025)[]{}; \draw[very thick, red] (0,0)--(0.3,0) }}
\DeclareRobustCommand\linelabelorange{\tikz{\node at (0.1,0.025)[]{}; \draw[very thick, orange] (0,0)--(0.3,0) }}

\DeclareRobustCommand\labelredarrow{\tikz{\node at (0.1,0.025)[]{}; \draw[-latex,very thick, red, dashed] (0,0)--(0.4,0) }}

\DeclareRobustCommand\labelbluearrow{\tikz{\node at (0.1,0.025)[]{}; \draw[-latex,very thick, blue] (0,0)--(0.4,0) }}
\DeclareRobustCommand\labelblackline{\tikz{\node at (0.1,0.025)[]{}; \draw[very thick, black] (0,0)--(0.4,0) }}
\DeclareRobustCommand\labelblackcircle{\tikz{
        \fill[black] (0.1,0.025) circle (2.5pt) }}
\DeclareRobustCommand\labelgrayarrow{\tikz{\node at (0.2,0.025)[]{}; \draw[-latex,very thick, gray, dashed] (0,0)--(0.4,0) }}

\title{Queue-aware Network Control Algorithm with a High Quantum Computing Readiness---Evaluated in Discrete-time Flow Simulator for Fat-Pipe Networks  

\iflatexml
\else
\thanks{This work has been performed in the framework of the CELTIC-NEXT EUREKA project AI-NET ANTILLAS (Project ID C2019/3-3), and it is partly funded by the German Federal Ministry of Education and Research (Project ID 16 KIS 1312). The author alone is responsible for the content of the paper.

The author thanks A. Kirstädter for fruitful discussions.}
\fi
}

\author{
\IEEEauthorblockN{Arthur Witt 
\iflatexml
\href{https://orcid.org/0000-0003-1180-1172}{[ORCID: 0000-0003-1180-1172]
}
\else
\orcidlink{0000-0003-1180-1172}
\fi}
\IEEEauthorblockA{\textit{Institute of Communication Networks and Computer Engineering, University of Stuttgart}, Stuttgart, Germany}
\IEEEauthorblockA{arthur.witt@ieee.org}
}

\maketitle
\begin{acronym}[WWW]
	\acro{DWDM}{dense wavelength division multiplexing}
	\acro{gRPC}{general-purpose remote procedure call}
	\acro{ILP}{integer linear program}
	\acro{IP}{internet protocol}
	\acro{NP}{non-deterministic polynomial}
	\acro{OTN}{optical transport network}
	\acro{OXC}{optical cross connects}
	\acro{QA}{quantum annealer}
	\acro{QC}{quantum computing}
	\acro{QUBO}{quadratic unconstrained binary optimization}
	\acro{SDN}{software-defined networking}
\end{acronym}

\begin{abstract} The emerging technology of quantum computing has the potential to change the way how problems will be solved in the future. This work presents a centralized network control algorithm executable on already existing quantum computer which are based on the principle of quantum annealing like the D-Wave Advantage\texttrademark. We introduce a resource reoccupation algorithm for traffic engineering in wide-area networks. The proposed optimization algorithm changes traffic steering and resource allocation in case of overloaded transceivers. Settings of active components like fiber amplifiers and transceivers are not changed for the reason of stability. This algorithm is beneficial in situations when the network traffic is fluctuating in time scales of seconds or spontaneous bursts occur. Further, we developed a discrete-time flow simulator to study the algorithm's performance in wide-area networks. Our network simulator considers backlog and loss modeling of buffered transmission lines. Concurring flows are handled equally in case of a backlog. 

This work provides an ILP-based network configuring algorithm that is applicable on quantum annealing computers. 
We showcase, that traffic losses can be reduced significantly by a factor of 2 if a resource reoccupation algorithm is applied in a network with bursty traffic. As resources are used more efficiently by reoccupation in heavy load situations, overprovisioning of networks can be reduced. Thus, this new form of network operation leads toward a zero-margin network. 
We show that our newly introduced network simulator enables analyses of short-time effects like buffering within fat-pipe networks. 
As the calculation of network configurations in real-sized networks is typically time-consuming, quantum computing can enable the proposed network configuration algorithm for application in real-sized wide-area networks. 
\end{abstract}


%
%
%

\iflatexml
\textbf{Copyright Notice}\\
\noindent\copyright 2024 IEEE. Personal use of this material is permitted. Permission from IEEE must be obtained for all other uses, in any current or future media, including reprinting/republishing this material for advertising or promotional purposes, creating new collective works, for resale or redistribution to servers or lists, or reuse of any copyrighted component of this work in other works.

\textbf{Acknowledgment}\\
This work has been performed in the framework of the CELTIC-NEXT EUREKA project AI-NET ANTILLAS (Project ID C2019/3-3), and it is partly funded by the German Federal Ministry of Education and Research (Project ID 16 KIS 1312). The author alone is responsible for the content of the paper.

The author thanks A. Kirstädter for fruitful discussions.
\else
\begin{textblock}{14}(-0.5,3.8)
	\fbox{
		\begin{minipage}{18.35cm}
			\noindent\copyright 2024 IEEE. Personal use of this material is permitted. Permission from IEEE must be obtained for all other uses, in any current or future media, including reprinting/republishing this material for advertising or promotional purposes, creating new collective works, for resale or redistribution to servers or lists, or reuse of any copyrighted component of this work in other works.
		\end{minipage}\hspace{1mm}}
\end{textblock}
\fi

\section{Introduction}
\label{sec:introduction}
\subsection{Motivation}\enlargethispage{0mm}
Optical transport networks and the involved technologies are under an ongoing redesign. The paradigm of software-defined networking (SDN) enables a centralized network control. Protocols like gRPC for the transport of telemetry data, and interfaces enabling the management of network components like Netconf and RESTconf, are essential parts of an SDN network control architecture. The optimization of network resources enables a low cost network operation---crucial for network operators and their customers. Some of the required optimization tasks are describable as combinatorial problems with a \acf{NP}-hard complexity. 

\begin{figure}[!h]
	\vspace{-1mm}
    \includegraphics[trim= 0.8cm 5.0cm 0.75cm 10.5cm, clip, width=\linewidth]{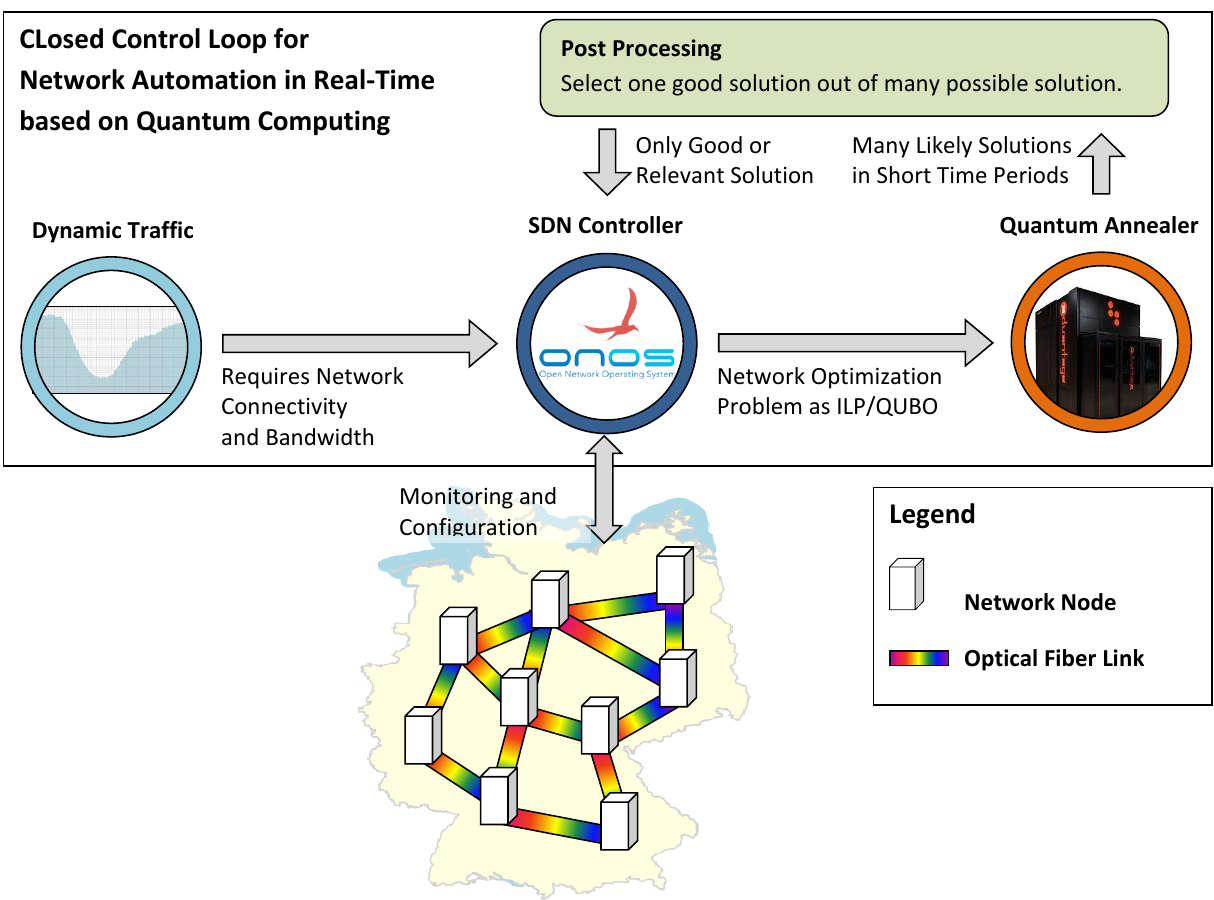}
    \caption{Conceptional architecture of SDN-based network control for wide-area networks with a quantum annealer as optimizer.}
    \label{fig:concept}
\end{figure}

Quantum computing has the potential to speed-up the computation of NP-hard problems in a polynomial to exponential order. Previous studies \cite{bib:QW_Paper}, \cite{bib:ilp_qa_ml} and others like \cite{bib:QAasILP} showed that integer linear programs can be solved on quantum annealer like the D-Wave Advantage\texttrademark. Particularly \cite{bib:QW_Paper} indicates that resources in optical networks can be optimized with quantum annealing. The network resource optimization problem modeled as ILP is transfered to a quadratic unconstrained binary optimization (QUBO) problem which is finally solvable on the quantum annealer. This problem mapping algorithm was firstly introduced in \cite{bib:QAasILP}. The study of \cite{bib:MIPonBosonicQC} shows approaches to solve mixed-integer programs with boson-sampling quantum computing. Thus, various quantum computing approaches are currently under investigation and ILP solving algorithms are an essential class of algorithms.  Differing to \cite{bib:QW_Paper} the current work proposes a modified ILP to showcase the benefit of ultra-fast network reconfiguration, applicable in the sub-second regime. 
The overall concept at a glance is depicted in Fig.~\ref{fig:concept}.\enlargethispage{1mm}

\subsection{Objectives}
The current work presents a novel network control algorithm, with the following properties:
\begin{itemize}
\item It takes advantages of a centralized perspective on the states of network equipment enabled by \ac{SDN}.
\item It considers, that network traffic volumes are normal distributed on time scales of seconds or less as observed in \cite{bib:Traffic_Stanford}. A further study \cite{bib:Traffic_Long_Term_Study} shows, that traffic volumes are log-normal distributed. However, they used normal and log-normal distributions for traffic modeling. Even if the log-normal model fits the traffic best, normal distributions show a remarkable fitting level as well. Thus, we assume that a normal-distribution can be used for a valid traffic rate modeling.
\item It is modeled in the form of an \acf{ILP} that can be solved on current available quantum annealing computers like the D-Wave Advantage\TM 5.2/5.3, which was partly proven by experiment in \cite{bib:QW_Paper}. It might be also solvable on a quantum computer like the coherent Ising machine with 100 thousand qubits presented in \cite{bib:100kqubit}.
\item It enables a two-fold optimization method. On larger time-intervals, e.g., in the range of several minutes, the network is reconfigured by activation of installed network hardware based on a average network traffic volume and a provisioning factor $\pi$. This includes also the traffic engineering, i.e., selection of routes and traffic migration.
As varying traffic volumes may cause the filling of network buffers within time scales in the order of seconds, a queue-length aware optimization is introduced to adapt the traffic engineering without changing the activity status of network hardware.
\end{itemize}  
Further, we provide a model of the optical transport network layer for chunk-wise flow simulations that was inspired by \cite{bib:FlowSim}.
Finally, the algorithm's performance is studied by time-discrete simulations based on the proposed network model.

\section{Methods and Materials}

\subsection{Network Architecture}
We consider an IP-over-DWDM network architecture, which is typically used in wide-area or nationwide networks. Those networks are composed of nodes $v\in V$ and links which are structured according to a meshed graph topology. For redundancy purposes each node is connected by at least two links with other nodes of the network.

Network nodes are two-layered. The upper layer is equipped with IP/MPLS routers and signals are processed electrically. The lower optical layer consists of optical cross connects that allow a wavelength-specific switching of optical signals and their forwarding. Signals between both layers are transformed by transceivers. They allow transmission and reception for flexible configuration, but used mostly unidirectional as configured. Since signal processing within transceivers is energy-intensive, we strive to minimize the consumption during network operation.

Bidirectional network links are realized by at least two optical fibers, which are used only unidirectional to avoid signal distortion. Fiber amplifiers are equipped (every 80 km) along the fibers to increase the optical reach of transceivers. The fiber's capacity is shared by dense wavelength division multiplexing, which is realized by optical signal multiplexing within a spectral grid of 50 GHz wide bands. 

We further consider that the network is configured by a centralized controller according to the \acf{SDN} paradigm. The gRPC protocol is used to push telemetry data from network devices to the controller. It can be assumed that small data portions in the order of multiple bytes each are send every \SI{50}{ms} to \SI{500}{ms} to keep information about the network up to date at the controller site. Protocols like NETCONF and RESTconf are required for configuring network devices. It can be considered that new configuration data per device are in the size of some mega bytes. There application may take some time, e.g. \SI{50}{ms}, as configurations are checked first for there consistency before application.
In our study we will consider the \ac{SDN} protocol activities only in form of fixed time delays for signaling and device configuration.
Essential part of the controller is a decision engine that performs the network optimization. We define two \acp{ILP} in the sections \ref{sec:ILP_constants} to \ref{sec:ILP_LT}. Both \acp{ILP} can be executed on a quantum annealer like the D-Wave Advantage\texttrademark. We performed our \ac{ILP}-based studies with a classical \ac{ILP}-solver.

\subsection{Traffic Model}
\label{sec:traffic_model}

Typical traffic rates follow a normal distribution within short time scales as shown in \cite{bib:Traffic_Stanford}. We have further to assume, that spontaneous data exchange, e.g. between data center, causes an unpredictable traffic burst. Thus, we synthesize the network traffic as follows:
\begin{enumerate}
\item Burst events appear in random intervals of 
$T_B\sim\mathcal{NEXP}(\lambda/|D|)$ with a negative exponential distribution. $\lambda$ defines the density of burst events per second. The normalization with the number of demands $|D|$ in the network allows one to compare similar situations in networks with different sizes.
\item We randomly select one demand $d\in D$ according to a uniform distribution, that will change the traffic volume due to a burst event.
\item Traffic volumes for the initialization of all demands and in case of a burst event are chosen from a normal distribution with a high variance, e.g. $\mu_\mathrm{ST}\sim \mathcal{N}(\mu_\mathrm{B}, \sigma_\mathrm{B})$. This value is a short term mean value and will be used to define the  short time traffic profile in the next step.
\item Traffic volumes for short time traffic changes are chosen frequently (every $\Tsim$) from a normal distribution with a rather small variance and the short term mean value from above, i.e. $h_d \sim\mathcal{N}(\mu_\mathrm{ST},\sigma_\mathrm{ST})$
\end{enumerate}

\subsection{Model of Parallel Circuits with Queuing}
\label{sec:circ_queue_model}
\begin{figure}
    \includegraphics{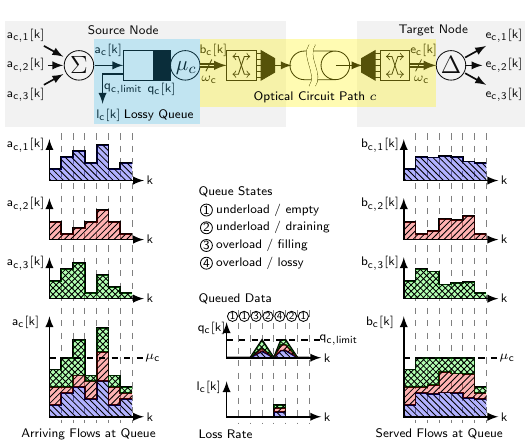}
    \caption{Time discrete model of parallel circuits along optical circuit path c for aggregated traffic flows. Overloads are handled by lossy buffering, realized with a time discrete queue model.}
    \label{fig:queue}
\end{figure}

We define a circuit path $c$ as an abstract transmission section that connects a source and target node optical-wise via fiber links. Circuit paths can use a single fiber on a direct way or by bypassing of intermediate nodes in the optical cross connect. It is realized by a number of equivalent circuits, encountered by $w_c$. All circuits have the same transmission rate of $\xi$, which ends-up in the combined service rate 
\begin{equation}
    \mu_c = \xi\cdot w_c\,.
\end{equation}
The length of a circuit path is limited by the transceiver-specific optical reach. \figref{\ref{fig:queue}} shows a typical transmission section that consist a) of a source node where multiple individual flows are aggregated and queued, b) a circuit path with parallel circuits connecting the source and target node, and finally c) a target node where the individual flows can be differentiated. The figure also depicts, how time-discrete flows are handled before, inside and after the queue. From hereon, we consider a time-discrete queuing and transport delay. 

The time delay for signal propagation along a circuit path $c$ with distance $D_c$ is simply expressed as integer multiple of the discretization interval $\Tsim$. Thus, the discrete propagation delay $\tau_c \Tsim$ is based on the integer factor
\begin{equation}
    \tau_c = \left\lceil v_p\cdot\frac{D_c\, \SI{}{[km]}}{T\, \SI{}{ [\text{\textmu}s] }}\right\rceil\,;
\end{equation}
$v_p$ is the considered propagation velocity of \SI{5}{\micro s /km}.

We consider that individual flows $a_{c,i}$, i.e. demands with different pairs of source and target nodes are assigned to the circuit path $c$ for transmission. We obtain a total data rate,
\begin{equation}
    a_c[k]=\sum_i a_{c,i}[k]\,,
\end{equation}
arriving at the queue of circuit path $c$ for transmission at the discrete time step $t_k=k\Tsim, k\in\N, k\geq 0$ with discretization interval $\Tsim$. For a well approximation of traffic volumes and resulting network metrics like delay, $\Tsim$ should be chosen small, i.e., in the order of \SI{0.1}{ms} or even below.

In cases where the arriving rates are higher than the transmission capacity of a circuit, data are queued. The queue, assigned to a circuit $c$, is characterized by a maximal storage volume,
\begin{equation}
q_{c,\mathrm{limit}}[k]=\mu_c[k]\cdot q_\mathrm{ratio} \cdot 1\mathrm{s}\,.
\end{equation}

The modeling of queuing is defined by four distinct steps to differentiate various aspects of queuing like draining and filling of the queue, queue losses and data transport without queuing. Within this steps a general view on total values is used, i.e., information about individual flows are not required. A further step is given to determine the flow-dependent values for the transmission rate, loss rate and queue length based on the prior calculated total values.

\textbf{1) Draining}\\
Assuming that the queue at circuit $c$ is already filled with data amount $q_c[k-1]\geq 0$, the queued backlog will drain in the next simulation step with a rate of
\begin{equation}
    x_{c,\mathrm{drain}}[k]=\min\{\mu_c[k],\, q_c[k-1]/\Tsim\}\,.
\end{equation}

\textbf{2) Non-queued Transmission}\\
Based on the arriving data amounts a portion of it can be transported without queuing. This rate is given by 
\begin{equation}
    x_{c,\mathrm{nqt}}[k]=\min\{a_c[k],\, \max\{0,\, \mu_c[k]-q_c[k-1]/\Tsim\}\}\,.
\end{equation}

\textbf{3) Filling the Queue}\\
The available space for data storage in the queue is defined as
\iflatexml
\begin{equation}
	q_{c,\mathrm{free}}[k]=\min\{q_{c,\mathrm{limit}}[k],\,
	\max\{0,\, q_{c,\mathrm{limit}}[k] - q_c[k-1]\} + x_{c,\mathrm{drain}}[k]\cdot\Tsim\}\,.
\end{equation}
\else
\begin{align}
    q_{c,\mathrm{free}}[k]=&\min\{q_{c,\mathrm{limit}}[k],\,
    \max\{0,\, q_{c,\mathrm{limit}}[k] - q_c[k-1]\} \nonumber\\
    &+ x_{c,\mathrm{drain}}[k]\cdot\Tsim\}\,.
\end{align}
\fi
Due to a traffic overload of 
\begin{equation}
    x_{c,\mathrm{overload}}=\max\{a_c[k]- x_{c,\mathrm{nqt}}[k],\, 0\}\,,
\end{equation}
the queue is filled with the filling rate,
\begin{equation}
    x_{c,\mathrm{fill}}[k]=\min\{x_{c,\mathrm{overload}},\, q_{c,\mathrm{free}}[k]/\Tsim\}\,.
\end{equation}

\textbf{4) Loss Rate}\\
The total loss rate per circuit can be given by
\begin{equation}
l_c[k]=\max\{x_{c,\mathrm{overload}}[k]-x_{c,\mathrm{fill}}[k], 0\}\,.
\end{equation}

\textbf{5) Flow-dependent Values and Final Steps}\\
A flow-dependent loss rate can be defined as
\begin{equation}
    l_{c,i}[k]=\begin{cases}
        \frac{a_{c,i}[k]}{a_{c}[k]}\cdot l_c[k]&\text{if}\quad a_c[k]>0\\
        0 &\text{otherwise}
    \end{cases}\,.
\end{equation}
Further, we define the traffic arrival share ratio, 
\begin{equation}
    \alpha_{c,i}[k]=\begin{cases}
        \min\left\{1,\,\frac{a_{c,i}[k]}{a_{c}[k]}\right\}&\text{if}\quad a_c[k]>0\\
        0&\text{otherwise}
    \end{cases}\,,
\end{equation}
and queuing share ratio, 
\begin{equation}
    \gamma_{c,i}[k]=\begin{cases}
        \min\left\{1,\,\frac{q_{c,i}[k-1]}{q_{c}[k-1]}\right\}&\text{if}\quad q_c[k-1]>0\\
        0&\text{otherwise}
    \end{cases}\,.
\end{equation}
Both ratios are limited to the interval $[0,1]$ to provide the required numerical stability. 
Thus, we can give the individual transmission rates, 
\begin{equation}
    b_{c,i}[k]=  
    \alpha_{c,i}[k] x_{c,\mathrm{nqt}}[k]+\gamma_{c,i}[k]x_{c,\mathrm{drain}}[k]   \,,
\end{equation}
and the flow dependent data amounts stored in the queue,
\iflatexml
\begin{equation}
	q_{c,i}[k]=\max\{\Tsim \left(
	\alpha_{c,i}[k] x_{c,\mathrm{fill}}[k]-\gamma_{c,i}[k]x_{c,\mathrm{drain}}[k]\right)+q_{c,i}[k-1],\,0
	\}\,.
\end{equation}
\else
\begin{align}
    q_{c,i}[k]=&\max\{\Tsim \left(
    \alpha_{c,i}[k] x_{c,\mathrm{fill}}[k]-\gamma_{c,i}[k]x_{c,\mathrm{drain}}[k]\right)\nonumber\\
    &+q_{c,i}[k-1],\,0
    \}\,.
\end{align}
\fi
Finally, the total queue length can be given by 
\begin{equation}
    q_c[k] = \sum_i q_{c,i}[k]\,,
\end{equation} 
which is required for the next simulation step. An equivalent relation exists for the total transmission rate $b_c[k]$.

\subsection{Variables and Constants for ILP-based Optimization}
\label{sec:ILP_constants}
A set of possible transmission path configurations $T_d$ is generated prior optimization accordingly to a well-defined process as described in \cite{bib:QW_Paper}. Therefore, connectivity relevant network properties are considered by the constants $\rho_{c,t_d}$ and $\varphi_{v,c}$. A selected configuration $t_d \in T_d$ describes how the traffic flow of end-to-end demand $d$ is realized. This considers routing, optical bypassing of network nodes and the chain-building of multiple optical transmission sections. Each section is terminated by transceivers at source and target nodes $v$.

\vspace{2mm}\noindent\textit{Variables:}
\begin{itemize}
    \setlength\itemsep{-0.2em}
    \item $g_{t_d} \in \{0,1\}$: path selector, \ie, $g_{t_d}$ equals 1 if a transmission path for demand $d$ is realized by transmission path configuration $t_d\in T_d$.
    \item $w_c \in \N$: the number of active, parallel circuits on circuit path $c$.
    \item $q_\mathrm{max} \in \R$: expected maximal queue length within optimization interval $\Delta T$.
\end{itemize}
\textit{Constants:}
\begin{itemize}
    \setlength\itemsep{-0.2em}
    \item $\xi \in \R$: the data rate of a single optical circuit.
    \item $\eta_v \in \N$: the amount of transceivers installed at node $v$.
    \item $\rho_{c,t_d} \in \{0,1\}$: indicates whether circuit configuration $t_d$ uses circuit path $c$.
    \item $\varphi_{v,c} \in \{0,1\}$: indicates whether node $v$ is the source or target node of circuit path $c$.
    \item $h_d\in \mathbb{R}$: traffic volume of demand $d$.
    \item $\Delta T$: time interval for frequently applied optimization (especially for resource reoccupation).
    \item $q_c$: actual queue length while optimization
\end{itemize}

\subsection{Resource Provisioning (Long-term Optimization)}
\label{sec:ILP_LT}

The authors of \cite{bib:QW_Paper} introduced an \ac{ILP} for network resource allocation and demonstrate its applicability on quantum annealing computers. We will reuse this  proposed \ac{ILP} in the following to achieve an optimal resource provisioning, i.e., the activation of optical transceivers, with a long-term perspective in the order of minutes. This low frequent reconfiguration of optical equipment preserves the network's stability in the optical network layer.

\iflatexml
\fbox{
\begin{minipage}{\textwidth}
	\textbf{ILP --- Resource Allocation}
	
	\noindent\textit{Constraints}:
	\begin{align}
		\sum_{t_d \in T_d}g_{t_d} = 1 \hspace{3mm}&\forall d \in D
		\label{eq:ilp_lt_const_demands}\\
		-\xi w_c + \sum_{d \in D}\sum_{t_d\in T_d} \pi h_{d,\mathrm{avg}}\rho_{c,t_d}   g_{t_d} \le 0 \hspace{3mm}&\forall c \in C 
		\label{eq:ilp_lt_const_circuits}\\
		\sum_{c \in C} w_c \varphi_{v,c} \le \eta_v  \hspace{3mm}&\forall v \in v 
		\label{eq:ilp_lt_const_nodes}\\
		\text{with}\quad g_{t_d} \in \{0,1\},\quad
		w_c \geq 0\nonumber
	\end{align}
	\textit{Objective:}
	\begin{eqnarray}
		\sum_{c \in C} w_c\ \rightarrow\ \min.\label{eq:ilp_lt_objective}
	\end{eqnarray}
\end{minipage}}
\else
\begin{tcolorbox}
\textbf{ILP --- Resource Allocation}
	
\noindent\textit{Constraints}:
\begin{align}
    \sum_{t_d \in T_d}\hspace{-1mm}g_{t_d} = 1 \hspace{3mm}&\forall d \in D
    \label{eq:ilp_lt_const_demands}\\
    -\xi w_c + \hspace{-1mm}\sum_{d \in D}\sum_{t_d\in T_d} \hspace{-1mm}\pi h_{d,\mathrm{avg}}\rho_{c,t_d}   g_{t_d} \le 0 \hspace{3mm}&\forall c \in C 
    \label{eq:ilp_lt_const_circuits}\\
    \sum_{c \in C} w_c \varphi_{v,c} \le \eta_v  \hspace{3mm}&\forall v \in v 
    \label{eq:ilp_lt_const_nodes}\\
    \text{with}\quad g_{t_d} \in \{0,1\},\quad
    w_c \geq 0\nonumber
\end{align}
\textit{Objective:}
\begin{eqnarray}
    \sum_{c \in C} w_c\ \rightarrow\ \min.\label{eq:ilp_lt_objective}
\end{eqnarray}
\end{tcolorbox}
\fi

Multiple optical circuits are allocatable under consideration of a traffic grooming in the IP/MPLS layer and proper path selection. The path selection is performed by \eqref{eq:ilp_lt_const_demands}, whereas grooming and allocation of optical circuits are done by \eqref{eq:ilp_lt_const_circuits}. Finally, \eqref{eq:ilp_lt_const_nodes} considers the limitation of equipped optical transceivers at the network nodes. 

The resource allocation is based on the average demand values $h_{d,\mathrm{avg}}$ for $d\in D$. Overprovisioning of channel capacity can be enforced with a provisioning factor $\pi>1$, otherwise $\pi=1$. The objective function \eqref{eq:ilp_lt_objective} minimizes the total amount of optical circuits. Thus, the number of active transceivers and the total  power-consumption of the network is also reduced.

\subsection{Dynamic Resource Reoccupation based on Queue Lengths (Short-term Optimization)}
\label{sec:ILP_ST}
An essential part of the current work is the development of an \ac{ILP} for short-time resource reoccupation in volatile traffic load situations. Traffic bursts may cause a data backlog in traffic flows if available transmission capacity is already occupied. This is typically handled by buffering to avoid a loss of traffic. We track and evaluate the queue lengths $q_c$ per optical circuit path $c$ and use it as indicator within our \ac{ILP}. We expect a circuit-specific queue length variation of $\Delta q_c$ within reconfiguration interval $\Delta T$, defined by
\begin{equation}
    \Delta q_c=\frac{\Delta T}{w_c}\biggl(\underset{\text{capacity}}{\underbrace{-\xi w_c}} + \underset{\text{traffic load}}{\underbrace{\sum_{d \in D}\sum_{t_\in T_d}   h_d  \rho_{c,t_d}g_{t_d} }} \biggr)\,. 
    \label{eq:delta_queue}\\
\end{equation}

In case of an increasing queue length, resource reoccupation might be applied, if and only if an equivalent optical circuit path with spare capacity exists. This is enforced by the ILP's objective \eqref{eq:ilp_st_objective} as is minimizes the largest queue length $q_\mathrm{max}$ in the network. Remark: Queue lengths are normalized to the capacity per optical channel. The maximal queue length has to fulfill the condition
\begin{equation}
	q_c + \Delta q_c \le q_\mathrm{max} \quad \forall c \in C\,. 
	\label{eq:queue} 
\end{equation}
This intuitively understandable condition leads together with \eqref{eq:delta_queue} to the formulation of constraint \eqref{eq:ilp_st_const_circuits}. 
It is considered that already activated optical circuits are reused. Hence, $w_c$ is a constant in this context. Only traffic routing and grooming are changed by \eqref{eq:ilp_st_const_demands} and \eqref{eq:ilp_st_const_circuits}. Thus, a resource reoccupation applicable in very short time scales, i.e. below seconds, is realized that does not affect the stability of the optical network. 

\iflatexml
\fbox{
\begin{minipage}{\textwidth}
\textbf{ILP --- Resource Reoccupation}

\noindent\textit{Constraints}:
\begin{align}
	\sum_{t_d \in T_d} g_{t_d} = 1 \hspace{3mm}&\forall d \in D
	\label{eq:ilp_st_const_demands}\\
	\hspace{-2mm}\frac{\Delta T}{w_c}\hspace{-0.5mm} \sum_{d \in D}\sum_{t_d \in T_d} \hspace{-1mm}\rho_{c,t_d}  h_d  g_{t_d} -q_\mathrm{max}\le \xi \Delta T- q_c \hspace{3mm}&\forall c \in C 
	\label{eq:ilp_st_const_circuits}\\
	\text{with}\quad g_{t_d} \in \{0,1\},\quad
	q_\mathrm{max} \geq 0\nonumber
\end{align}
\textit{Objective:}
\begin{eqnarray}
	q_\mathrm{max} \rightarrow\ \min.\label{eq:ilp_st_objective}
\end{eqnarray}
\end{minipage}}		
\else
\begin{tcolorbox}
\textbf{ILP --- Resource Reoccupation}

\noindent\textit{Constraints}:
\begin{equation}
    \sum_{t_d \in T_d} g_{t_d} = 1 \hspace{3mm}\forall d \in D
    \label{eq:ilp_st_const_demands}\\
\end{equation}
\begin{equation}
    \hspace{-2mm}\frac{\Delta T}{w_c}\hspace{-0.5mm} \sum_{d \in D}\sum_{t_d \in T_d} \hspace{-1mm}\rho_{c,t_d}  h_d  g_{t_d} -q_\mathrm{max}\le \xi \Delta T- q_c \hspace{3mm}\forall c \in C 
    \label{eq:ilp_st_const_circuits}\\
\end{equation}
\begin{equation}
	\text{with}\quad g_{t_d} \in \{0,1\},\quad
	q_\mathrm{max} \geq 0\nonumber
\end{equation}
\textit{Objective:}
\begin{eqnarray}
    q_\mathrm{max} \rightarrow\ \min.\label{eq:ilp_st_objective}
\end{eqnarray}
\end{tcolorbox}
\fi

\section{Evaluation and Discussion}

\subsection{Remarks on the Computation of the ILPs}
During the last decades, ILPs for network optimization have been studied mostly for the purpose of static optimization as their calculation last several minutes to hours, see \cite{bib:ILP_Tornatore}. Therefore, ILPs are atypical for dynamic network optimization since heuristic methods may be faster. We propose the use of ILPs as they allow a precise and well understandable modeling which is not necessarily the case for a heuristic approach. 

It is shown in \cite{bib:QW_Paper} and \cite{bib:QAasILP} that ILPs can be solved on a quantum annealing hardware. Therefore, ILPs are reformulated as quadratic unconstrained binary optimization (QUBO) problems. They are further mapped to an Ising Hamiltonian which can be then embedded and solved on a quantum annealing hardware like the D-Wave Advantage\texttrademark. 
Quantum annealing is also known as adiabatic quantum computing. Embedded problems have to be sampled as the quantum annealing procedure follows the stochastically laws of quantum mechanics. Quantum computing in general holds the premise to solve particular problems with a quadratic to exponential speed-up compared to classical computation. Currently, there are still remaining hurdles that need to be resolved prior quantum annealing can be used to solve realistic ILPs in reasonable time, see \cite{bib:QW_Paper} and \cite{bib:ilp_qa_ml}. However, we postulate that the computation time for solving ILPs can be reduced by quantum computing within the near future.

The resource allocating ILP, see Sec.~\ref{sec:ILP_LT}, was already solved with the help of a quantum annealer in the study of \cite{bib:QW_Paper}. There, solutions could be found within \SI{50}{sec.}. It was further observed, that 0.2 to 11 feasible ILP solutions could be found within 1 Mio.\ samples.

The current study analyzes the impact of fast solvable ILPs on the task of network control--not the ILP solving approach with quantum annealing itself. Therefore, a classical ILP solver was used to avoid the additional overhead of controlling the quantum annealer.
 
\subsection{Simulation Scenario}
A small network with 4 nodes is used for the evaluation of the proposed algorithms. The corresponding topology is depicted in \figref{\ref{subfig-1:network}}. Figures \ref{subfig-2:network} and \ref{subfig-3:network} show specific circuit path configurations that will be discussed later in Sec.~\ref{sec:results}. 
\iflatexml
\begin{figure}[h]
	\subfloat[Network topology to evaluate the \acp{ILP} \label{subfig-1:network}]
	{\includegraphics[width=0.32\linewidth]{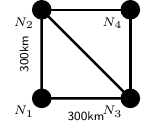}}
	\subfloat[Preferred traffic assignment\label{subfig-2:network}]
	{\includegraphics[width=0.32\linewidth]{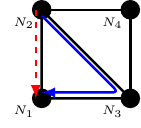}}
	\subfloat[Alternative traffic assignment\label{subfig-3:network}]
	{\includegraphics[width=0.32\linewidth]{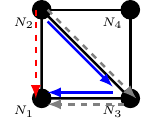}}
	\caption{(\labelblackcircle) Network nodes, (\labelblackline) fiber links, (\labelredarrow) network demand of interest, (\labelgrayarrow) network demands that share circuit paths with the demand of interest, (\labelbluearrow) circuit paths used to realize the demand of interest as traffic flow. 
	}
	\label{fig:topology}
\end{figure}
\else
\begin{figure}[h]
    \hspace{-1mm}
    \subfloat[Network topology to evaluate proposed \acp{ILP} \label{subfig-1:network}]
        {\hspace{-2mm}\includegraphics[width=0.33\linewidth]{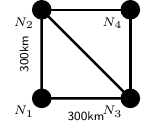}}
    \hspace{2mm}
    \subfloat[Preferred traffic assignment\label{subfig-2:network}]
        {\hspace{-2mm}\includegraphics[width=0.33\linewidth]{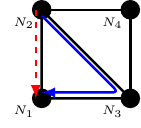}}
    \hspace{2mm}
    \subfloat[Alternative traffic assignment\label{subfig-3:network}]
        {\hspace{-2mm}\includegraphics[width=0.33\linewidth]{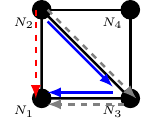}}
    \caption{(\labelblackcircle) Network nodes, (\labelblackline) fiber links, (\labelredarrow) network demand of interest, (\labelgrayarrow) network demands that share circuit paths with the demand of interest, (\labelbluearrow) circuit paths used to realize the demand of interest as traffic flow. 
    }
    \label{fig:topology}
\end{figure}
\fi

The network components are parametrized according to sections \ref{sec:circ_queue_model} and \ref{sec:ILP_constants} as follows. The channel capacity $\xi$ is set to \SI{100}{Gbit/s}. This corresponds to typical data rates in optical networks. E.g., the nominal service data rate for the often used OTU4 channels is around \SI{95}{Gbit/s} as defined in the \ac{OTN} standard \cite{bib:OTN_Standard}. 

We set $q_\mathrm{ratio}=0.05$ which corresponds to a maximal queue size of $q_{c,\mathrm{limit}} =\SI{5}{Gbit}$ if a single channel is used at circuit path $c$. Queues or buffers are typically sized in a way that a required bandwidth-delay product $\xi\cdot t_\mathrm{RTT}$ is ensured. For nation-wide networks round-trip times of 50 to \SI{100}{ms} are expectable. As queues are scaled proportionally to the circuit bandwidth, the round-trip time supported by the queue is defined as $q_\mathrm{ratio}\cdot\SI{1}{s}=\SI{50}{ms}$.
We limit the amount of installed transceivers per node to $\eta_v = 31$. 

\begin{figure}[h]
    \vspace{-4mm}
    \includegraphics[width=\linewidth]{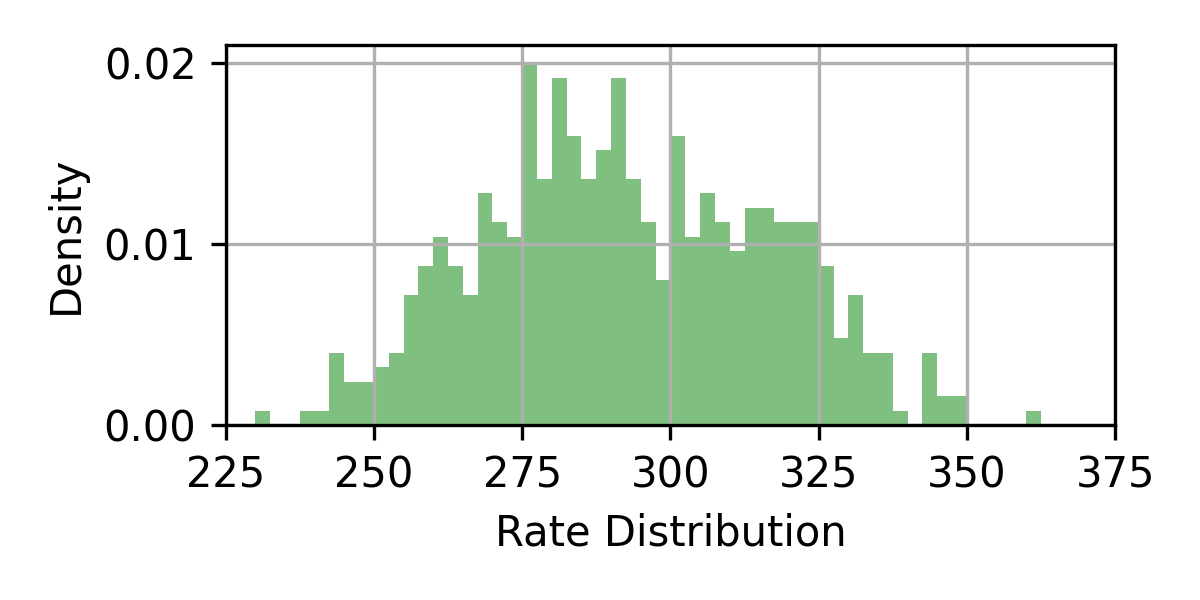}\vspace{-4mm}
    \caption{The data rate of a demand, cf. Fig.~\ref{fig:data_rate_bursts}, is normal distributed if traffic traces are analyzed over a long time duration. In short-time analyses the data rate shows a multi-modal distribution, which is caused by the traffic's burstiness.}
    \label{fig:data_rate_dist}\vspace{-5mm}
\end{figure}

According to our traffic model presented in \secref{\ref{sec:traffic_model}}, demands are generated with a mean value of $\mu_\mathrm{B}=\SI{290}{Gbit/s}$, and standard deviations for bursts  $\sigma_\mathrm{B}=\SI{30}{Gbit/s}$ and for short term fluctuations $\sigma_\mathrm{ST}=\SI{10}{Gbit/s}$. Burst events appear with a density of $\lambda=\SI{1}{s^{-1}}$. Fig.~\ref{fig:data_rate_bursts} shows an example of the modeled data rate for a particular traffic demand in the 4-node network. It is visible, that multiple burst occur during the shown time range of 5 seconds. Those bursts will cause the multi-modality of the mostly Gaussian data rate distribution, depicted in Fig.~\ref{fig:data_rate_dist}

\iflatexml
\begin{figure}[h]
	\includegraphics[width=\linewidth]{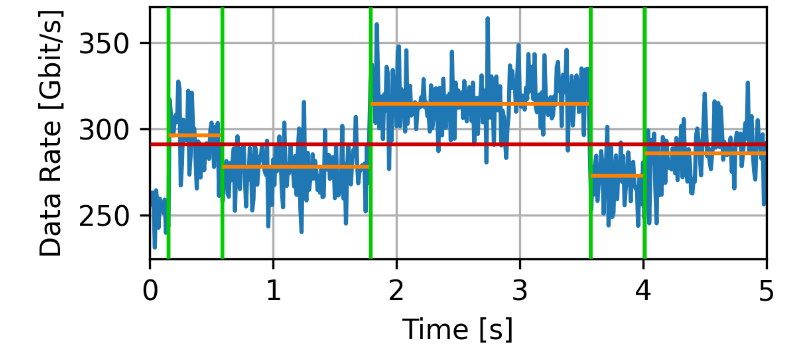}
	\caption{(\linelabelblue) Data rate of a traffic demand with a well-defined burstiness. (\linelabelgreen) Time instances of burst events with negative exponential distributed inter arrival times. (\linelabelorange) Mean value per burst, overlayed by short-term fluctuations. (\linelabelred) Long-term mean value staying constant for minutes or changing slowly.}
	\label{fig:data_rate_bursts}
\end{figure}
\else
\begin{figure}[h]
    \vspace{-4mm}
    \begin{tikzpicture}
        \node at(0,0)[text width=\linewidth]{\includegraphics[width=\linewidth]{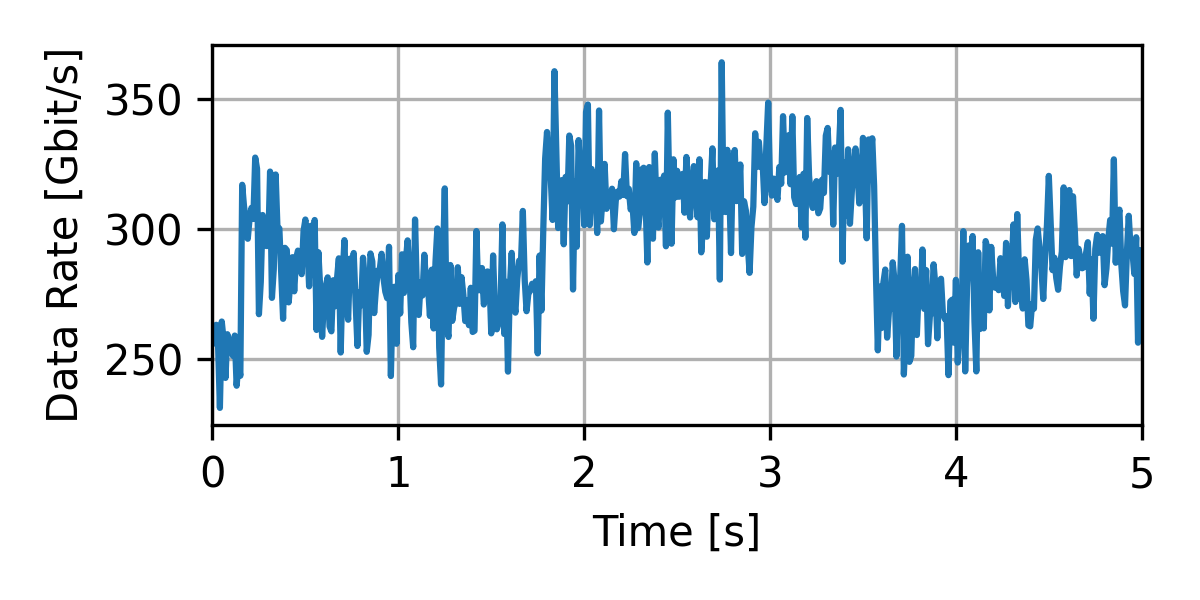}};
        \draw [very thick, green!80!black!100] (-2.65,-0.925)--(-2.65,1.875);
        \draw [very thick, orange] (-2.65, 0.45)--(-2.05,0.45);
        \draw [very thick, green!80!black!100] (-2.05,-0.925)--(-2.05,1.875);
        \draw [very thick, orange] (-2.05, 0.1)--(-0.4,0.1);
        \draw [very thick, green!80!black!100] (-0.4,-0.925)--(-0.4,1.875);  
        \draw [very thick, orange] (-0.4, 0.8)--(2.05,0.8);     
        \draw [very thick, green!80!black!100] (2.05,-0.925)--(2.05,1.875);
        \draw [very thick, orange] (2.05, -0.0)--(2.65,-0.0);
        \draw [very thick, green!80!black!100] (2.65,-0.925)--(2.65,1.875);
        \draw [very thick, orange] (2.65, 0.25)--(4,0.25);
        
        \draw [very thick, red!80!black!100] (-2.85,0.35)--(4,0.35);
    \end{tikzpicture}\vspace{-5mm}
    \caption{(\linelabelblue) Data rate of a traffic demand with a well-defined burstiness. (\linelabelgreen) Time instances of burst events with negative exponential distributed inter arrival times. (\linelabelorange) Mean value per burst, overlayed by short-term fluctuations. (\linelabelred) Long-term mean value staying constant for minutes or changing slowly.}
    \label{fig:data_rate_bursts}
\end{figure}
\fi

It is further considered, that a SDN controller is polling the states of network equipment every \SI{35}{ms} with a signaling delay of \SI{5}{ms}. The roll-out of new network configurations, that are obtained by the proposed optimization methods, is considered with a time delay of \SI{30}{ms}. Hereby, we are considering that RESTconf \emph{GET} or gRPC  operations are applied for status polling and HTTP-based RESTconf \emph{SET} operations are used to configure network devices. Typical latencies for SDN control and management protocol actions are studied experimentally in \cite{bib:sdn} and use are used to determine the proposed time delays.

\subsection{Functional Demonstration of Resource Reoccupation}
\label{sec:results}
In the following, two network operation modes are compared. In the \emph{resource allocation} mode, we will apply a network resource allocation as described in \secref{\ref{sec:ILP_LT}}. This mode considers the de-/activation of optical transceivers. Therefore, it is applied within intervals in the order of minutes to preserve the stability of the optical fiber system. For the evaluation we applied it just a single time at $t=0$ as a short time duration of \SI{5}{s} is analyzed only. For the evaluation of the \emph{resource reoccupation} mode we apply the optimization as described in \secref{\ref{sec:ILP_ST}} additionally to the resource allocation at $t=0$. The resource reoccupation is applied every \SI{100}{ms}. In both cases, a resource overprovisioning factor of $\pi=1.1$ is used. 

Figures \ref{subfig-1:sim1demand} and \ref{subfig-1:sim2demand} depict the requested data rates over time of all traffic demands in the network for the two applied network modes. It can be seen that the same traffic profiles are used for both network operation modes to provide a fair comparison. 

In case of the purely resource allocation, received data rates (see Fig.~\ref{subfig-2:sim1demand}) are mostly equivalent to the requested data rates (see Fig.~\ref{subfig-1:sim1demand}). Only demand $[N_2,N_1]$ shows a receive rate limitation within the time span between 2 and \SI{4}{s}, observable as homogeneously  colored area, indicating a receive rate of \SI{300}{Gbit/s}. As the requested rate of this demand is much higher within the particular time span, \SI{14.7}{Gbit} of data are lost as depicted in Fig.~\ref{subfig-3:sim1demand}. This graph shows data losses per demand cumulated over time. As a reference, each of the 12 traffic flows transports within the \SI{5}{s} a data volume of around \SI{1500}{Gbit}, such that a total relative loss of $0.8\permil$ is present.

\iflatexml
\begin{figure}
	\centering
	\subfloat[\label{subfig-1:sim1demand}]{\includegraphics[height=3.53cm, trim=4mm 0 0 0, clip]{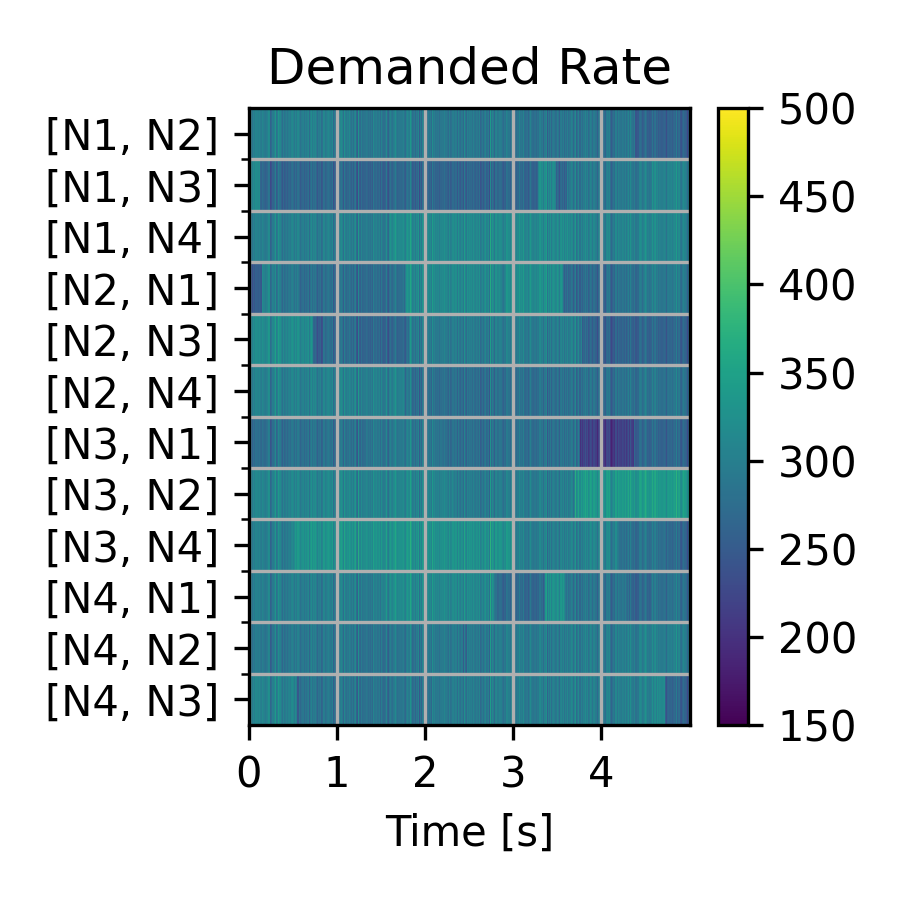}}
	\subfloat[\label{subfig-2:sim1demand}]{
		\includegraphics[height=3.53cm, trim=4mm 0 0 0, clip]{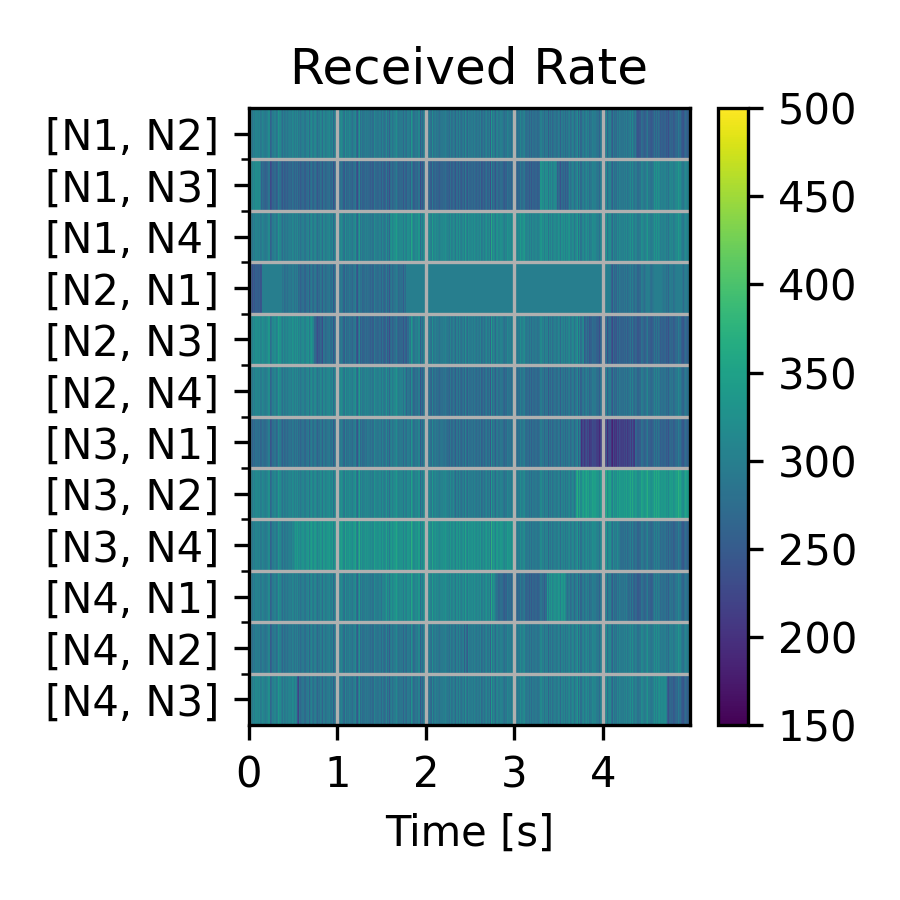}}
	\subfloat[\label{subfig-3:sim1demand}]{
		\includegraphics[height=3.48cm, trim=0 0 2mm 0, clip]{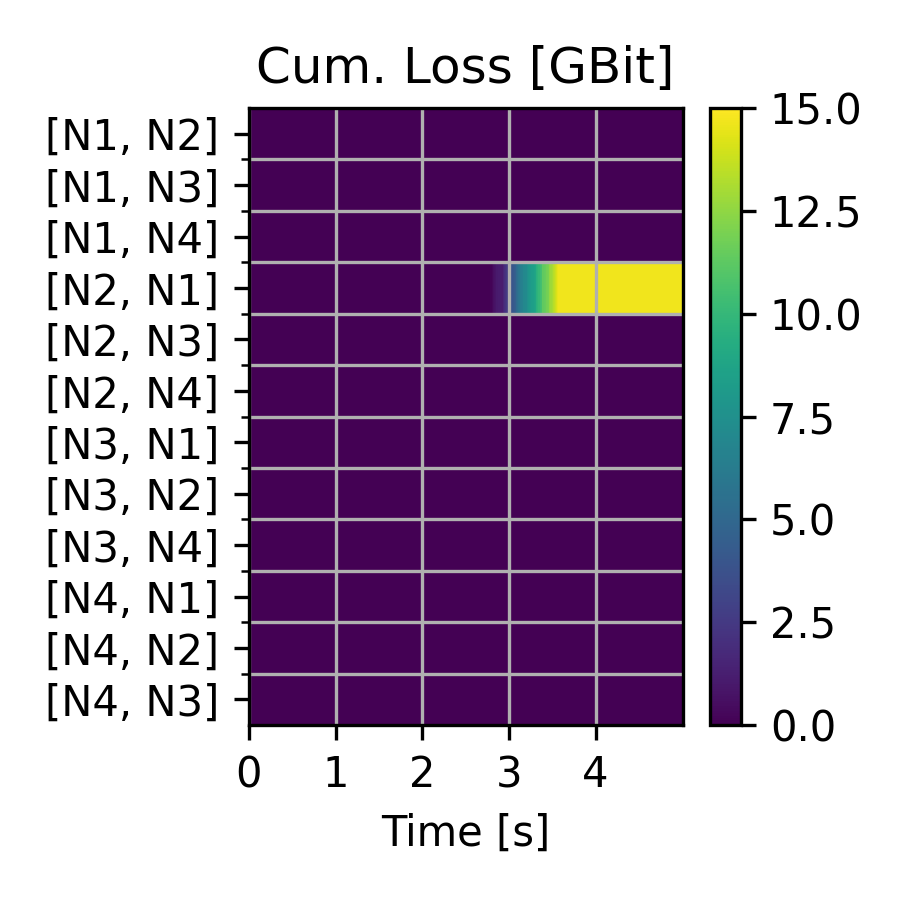}
	}	
	\caption{Data rates in \si{Gbit/s} of (a) network traffic demands at source nodes and (b) received flows at target nodes; (c) loss volumes per traffic flow in \si{Gbit} cumulated over time. The \emph{resource allocation} mode is applied.}
	\label{fig:sim1demand}
\end{figure}
\else
\begin{figure}
    \centering
    \vspace{-4mm}
    \subfloat[\label{subfig-1:sim1demand}]{\includegraphics[height=3.53cm, trim=4mm 0 1.6cm 0, clip]{sim_1_flow_act_cm.png}}
    \subfloat[\label{subfig-2:sim1demand}]{
    \includegraphics[height=3.53cm, trim=2cm 0 0 0, clip]{sim_1_flow_rec_cm.png}}
    \subfloat[\label{subfig-3:sim1demand}]{
    \includegraphics[height=3.48cm, trim=0 0 2mm 0, clip]{sim_1_loss_cm}
    }\vspace{-1mm}
    
    \caption{Data rates in \si{Gbit/s} of (a) network traffic demands at source nodes and (b) received flows at target nodes; (c) loss volumes per traffic flow in \si{Gbit} cumulated over time. The \emph{resource allocation} mode is applied.}
    \label{fig:sim1demand}
\end{figure}
\fi

\begin{figure}
    \centering
    \vspace{-4mm}
        \subfloat[\label{subfig-1:sim1circuit}]{%
           \includegraphics[width=0.49\linewidth]{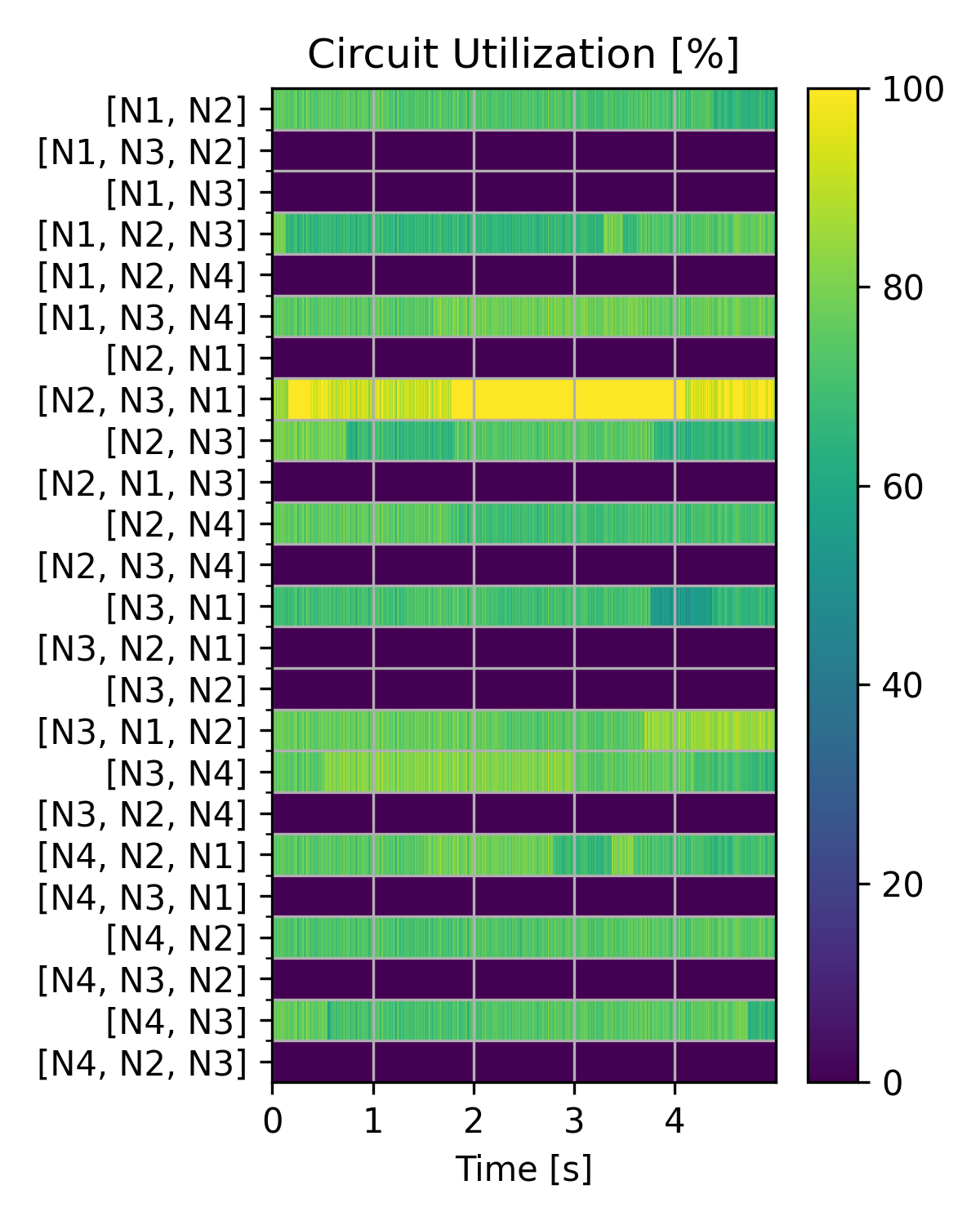}
        }
        \hspace{-2mm}
        \subfloat[\label{subfig-2:sim1circuit}]{%
            \includegraphics[width=0.49\linewidth]{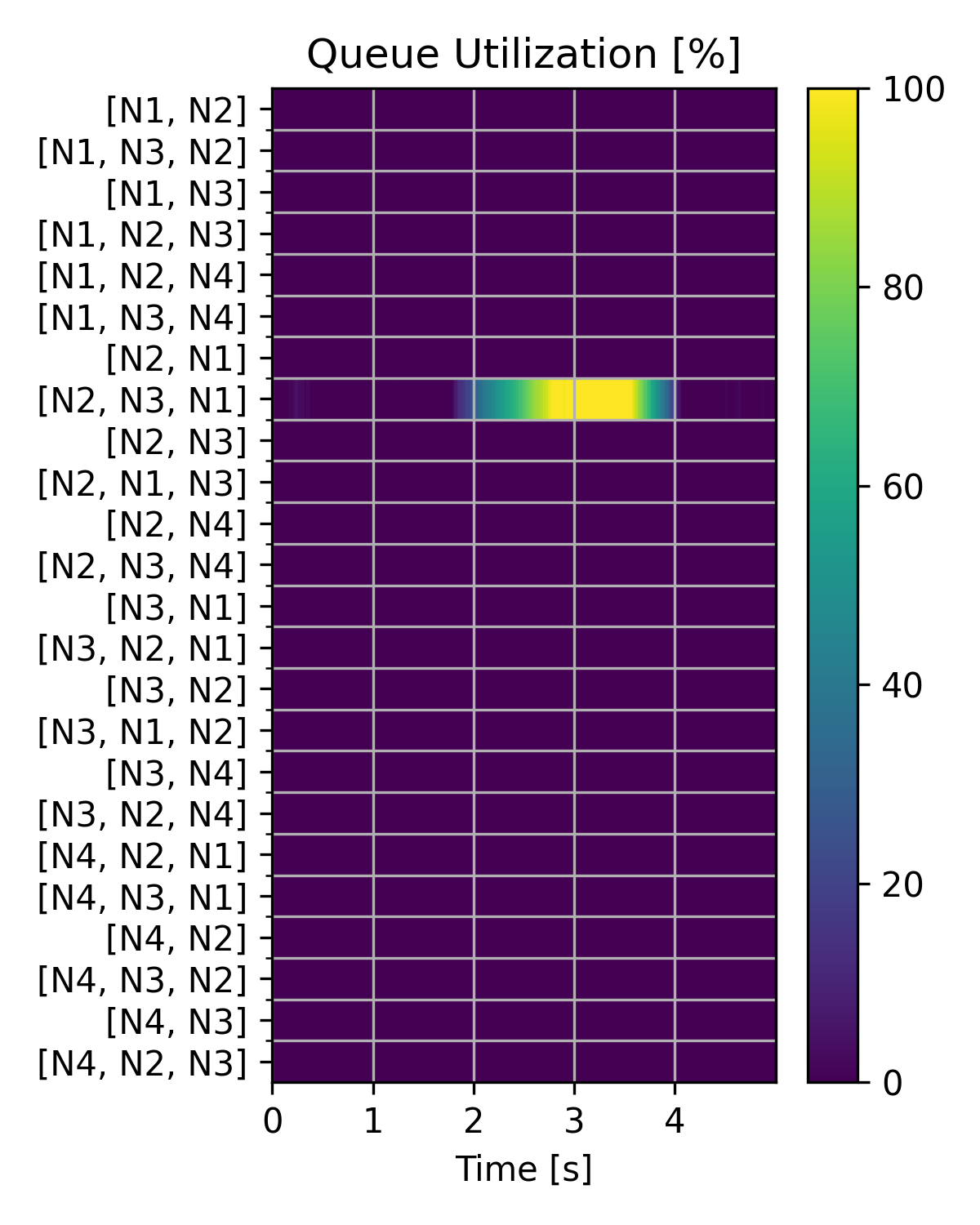} 
        }\vspace{-1mm}
        \caption{Utilization of (a) circuits and (b) assigned queues in percent. Values are normalized to the dynamic capacity per circuit and corresponding queue limits. The \emph{resource allocation} mode is applied.}
        \label{fig:sim1circuit}\vspace{-5mm}
\end{figure}

The analyses of the hardware utilization provide a better understanding of the obtained losses. Fig.~\ref{subfig-1:sim1circuit} depicts the utilization of circuit paths normalized to its capacity $\mu_c$ whereas Fig.~\ref{subfig-2:sim1circuit} shows the utilization of queues per circuit which is normalized to the queue limit $q_{c,\mathrm{limit}}$. It is observable, that only half of all circuit paths are used due to the selection obtained by the resource allocation optimization. Circuit path $[N_2,N_3,N_1]$ is highly loaded and its queue reached the maximal utilization at \SI{2.8}{s}. This causes finally the losses observed in Fig.~\ref{subfig-3:sim1demand}. 

\begin{figure}
    \centering
    \vspace{-4mm}
    \subfloat[\label{subfig-1:sim2circuit}]{
        \includegraphics[width=0.49\linewidth]{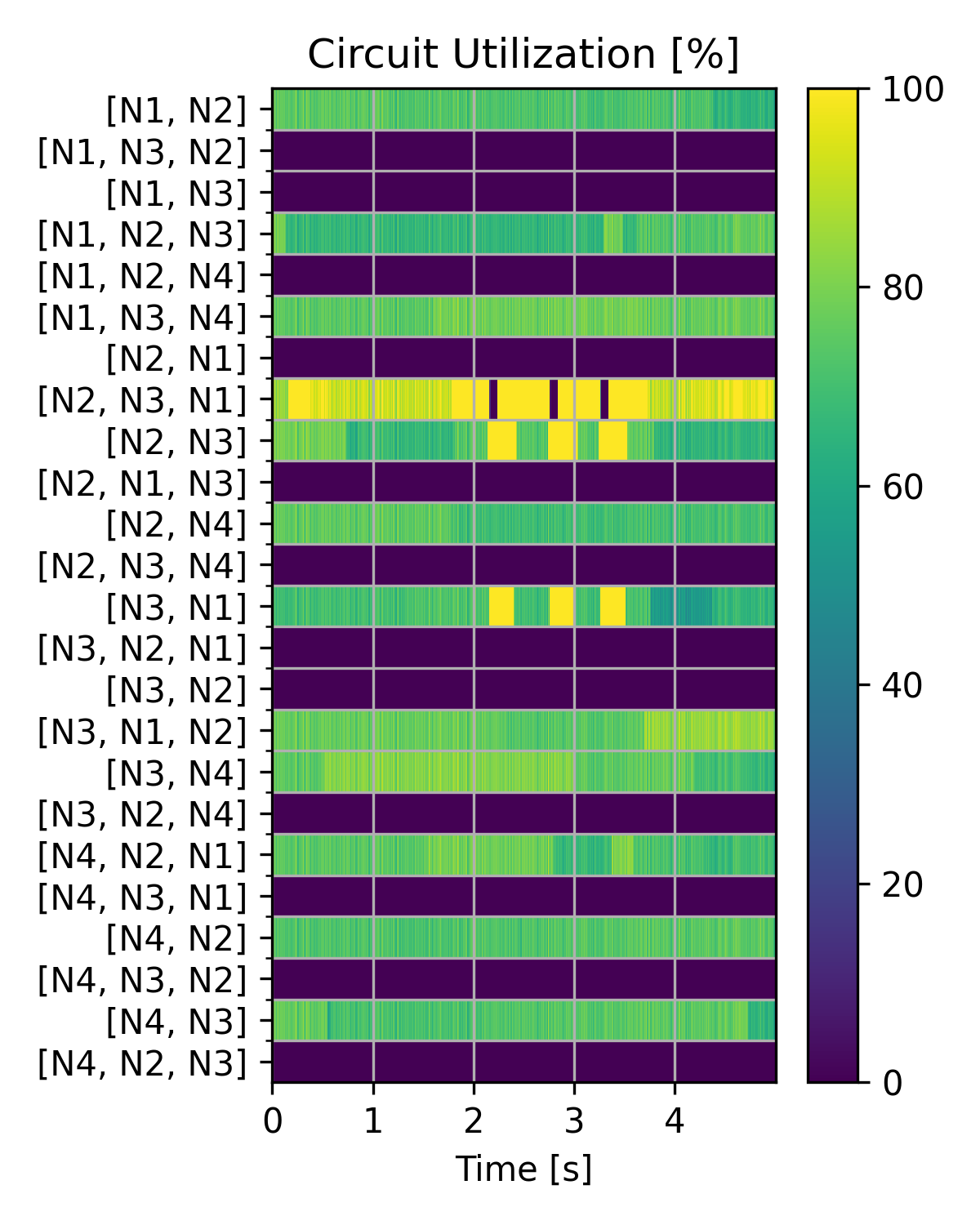}}
    \subfloat[\label{subfig-2:sim2circuit}]{
        \includegraphics[width=0.49\linewidth]{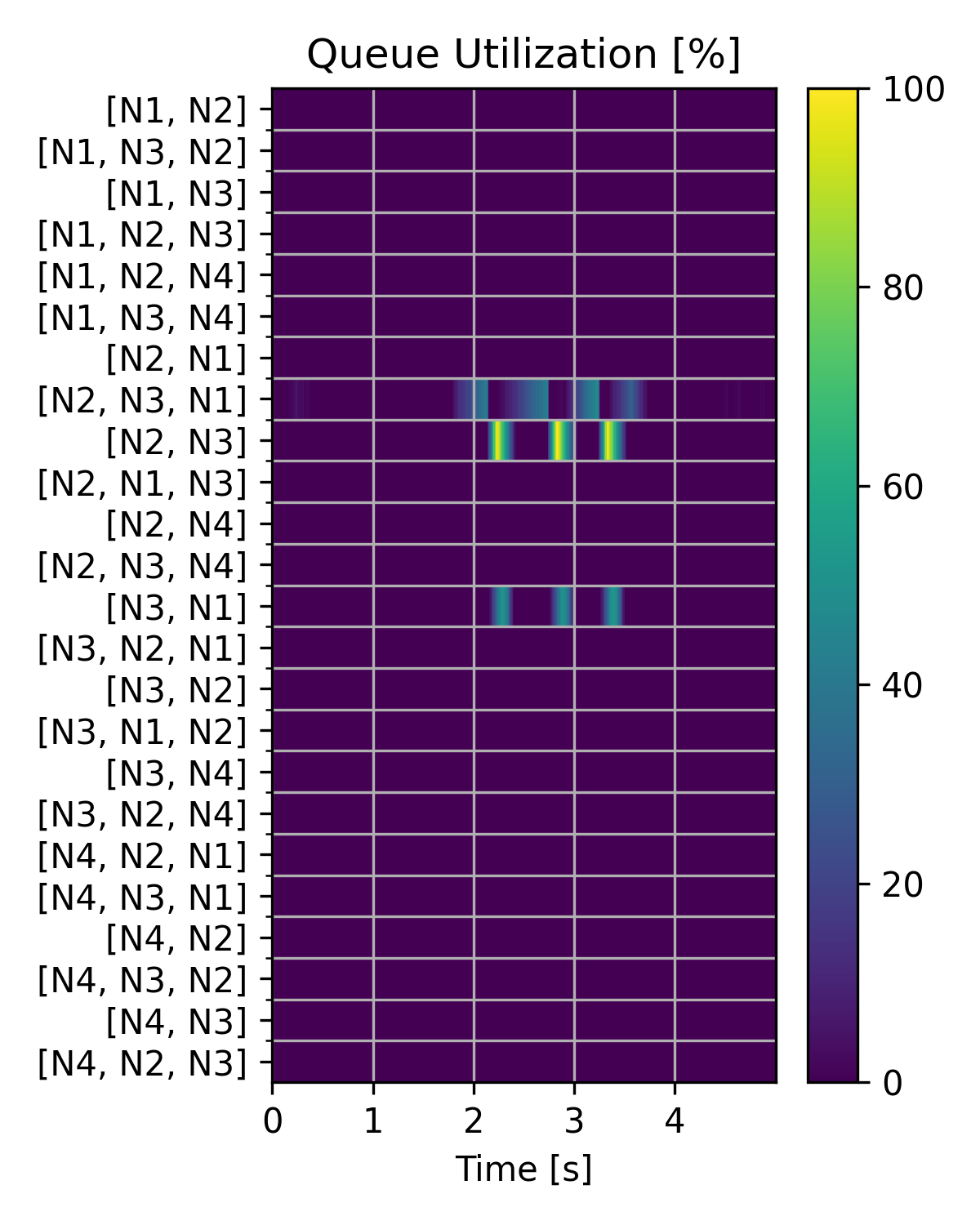}}\vspace{-2mm}
    \caption{Utilization of (a) circuits and (b) assigned queues in percent. Values are normalized to the dynamic capacity per circuit and corresponding queue limits. The \emph{resource reoccupation} mode is applied.}
    \label{fig:sim2circuit}\vspace{-3mm}
\end{figure}

In contrast to a pure resource allocation, the resource reoccupation mode avoids the overflow of buffers by offloading traffic to an alternative circuit path. 
As already observed in Fig.~\ref{fig:sim1demand}, demand $[N_2,N_1]$ is causing the lossy overload situation. In the resource allocation mode a corresponding flow is realized by utilization of circuit path $[N_2,N_3,N_1]$ as depicted in Fig.~\ref{subfig-2:network}. Resource reoccupation allows the offloading to the alternative circuit path sequence $[N_2,N_3], [N_3,N_1]$ as depicted in Fig.~\ref{subfig-3:network}. Therefore, the capacity per circuit path of this sequence has to be shared with demands $[N_2,N_3]$ and $[N_3,N_1]$. Fig.~\ref{subfig-1:sim2circuit} shows the circuit utilization for the resource reoccupation mode. The offloading of demand $[N_2, N_1]$ to the alternative circuit path sequence is used for three times within the time span of \SI{2}{s} to \SI{3.5}{s}. The queue utilization for the reoccupation mode (Fig.~\ref{subfig-2:sim2circuit}) indicates, that the queues are less loaded and occupied within a lower time duration compared to the resource allocation mode (Fig.~\ref{subfig-2:sim1circuit}). The load offloading activities causes an increased burstiness at the receiving target nodes (compare Fig.~\ref{subfig-2:sim2demand} with Fig.~\ref{subfig-2:sim1demand}). Hence, traffic at the wide-area network's target nodes will be distributed over multiple flows to reach different end-nodes in access networks, the burstiness is also distributed and less significant per flow. Finally, the resource reoccupation mode reduces the cumulated loss of demand $[N_2,N_1]$ to \SI{5.4}{Gbit}. In addition we obtain a loss for demand $[N_2, N_3]$ of \SI{2.2}{Gbit}, which is caused by the concurring flows at circuit path $[N_2, N_3]$. Those losses are depicted in Fig.~\ref{subfig-3:sim2demand}. In comparison to Fig.~\ref{subfig-3:sim1demand}, an overall loss reduction by a factor around 2 is obtainable.

\iflatexml
\begin{figure}
	\centering
	\subfloat[\label{subfig-1:sim2demand}]{
		\includegraphics[height=3.53cm, trim=4mm 0 0 0, clip]{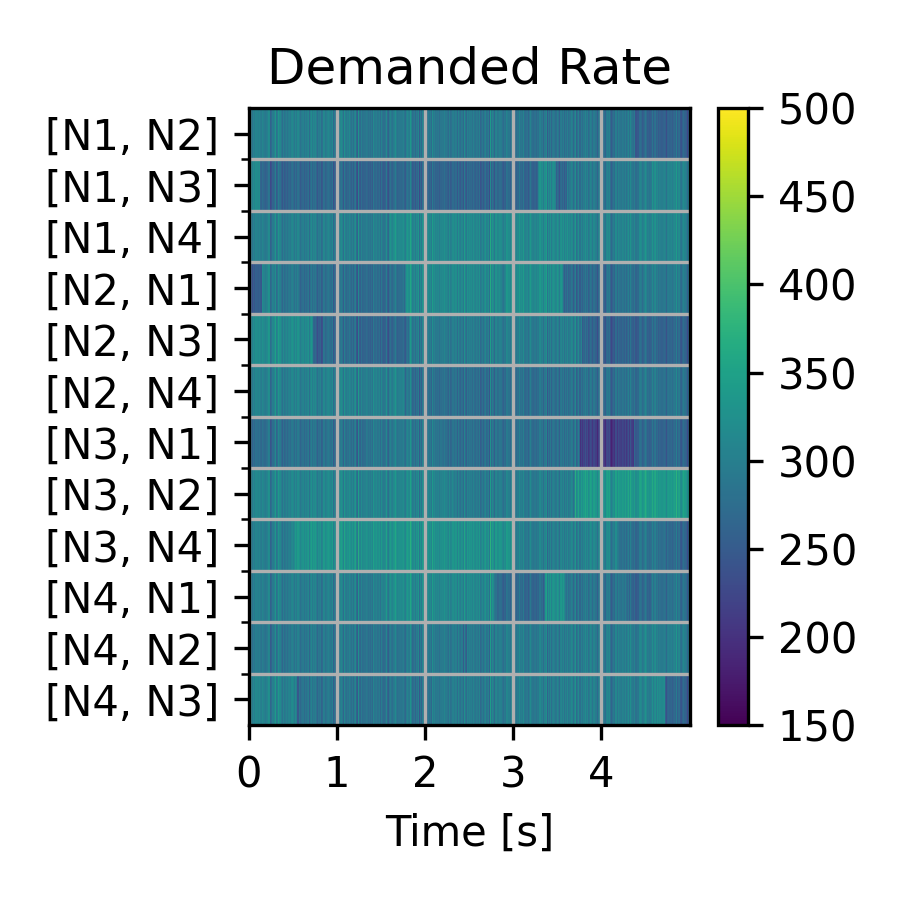}}
	\subfloat[\label{subfig-2:sim2demand}]{
		\includegraphics[height=3.53cm, trim=4mm 0 0 0, clip]{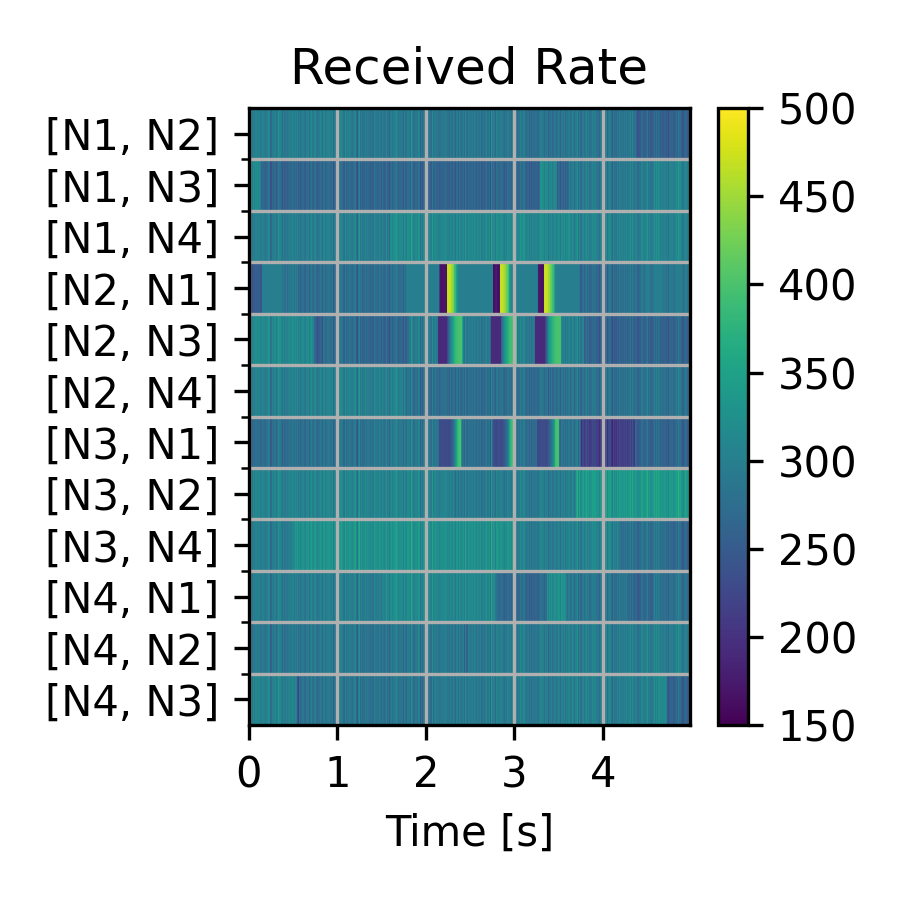}}
	\subfloat[\label{subfig-3:sim2demand}]{
		\includegraphics[height=3.48cm, trim=0 0 2mm 0, clip]{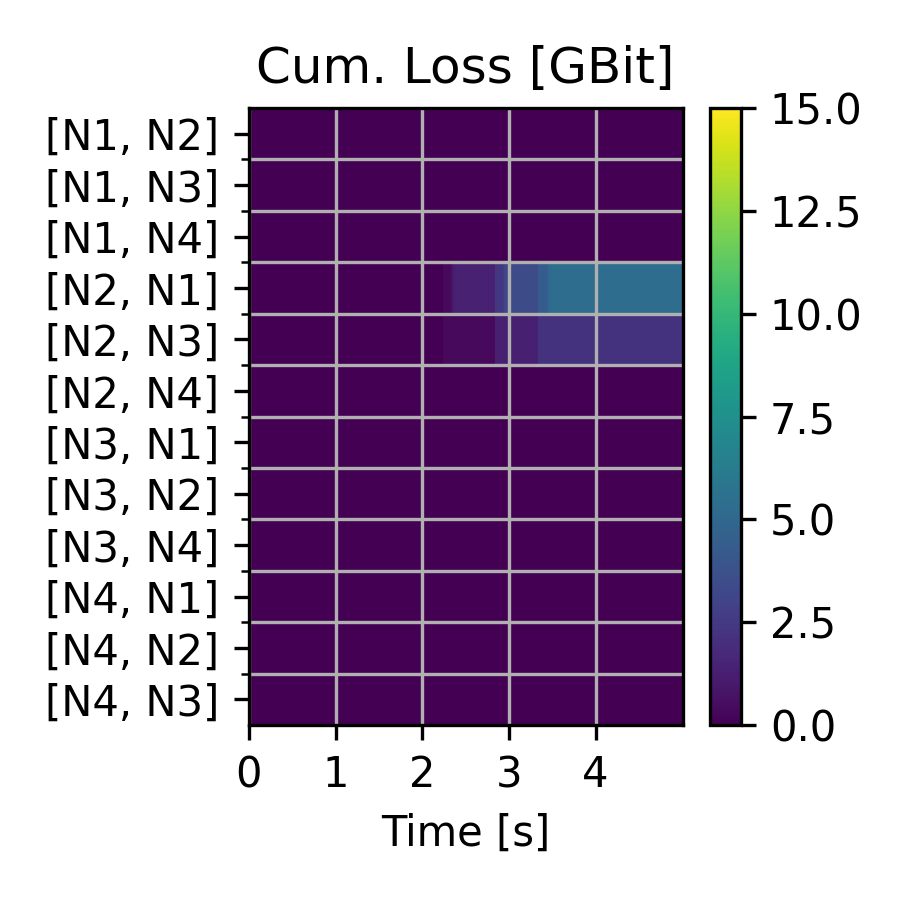}}
	\caption{Data rates in \si{Gbit/s} of (a) network traffic demands at source nodes and (b) received flows at target nodes; (c) loss volumes per traffic flow in \si{Gbit} cumulated over time. The \emph{resource reoccupation} mode is applied.}
	\label{fig:sim2demand}
\end{figure}
\else
\begin{figure}
    \centering
    \vspace{-4mm}
    \subfloat[\label{subfig-1:sim2demand}]{
        \includegraphics[height=3.53cm, trim=4mm 0 1.6cm 0, clip]{sim_2_flow_act_cm}}
    \subfloat[\label{subfig-2:sim2demand}]{
        \includegraphics[height=3.53cm, trim=2cm 0 0 0, clip]{sim_2_flow_rec_cm}}
    \subfloat[\label{subfig-3:sim2demand}]{
        \includegraphics[height=3.48cm, trim=0 0 2mm 0, clip]{sim_2_loss_cm}}\vspace{-2mm}
    \caption{Data rates in \si{Gbit/s} of (a) network traffic demands at source nodes and (b) received flows at target nodes; (c) loss volumes per traffic flow in \si{Gbit} cumulated over time. The \emph{resource reoccupation} mode is applied.}
    \label{fig:sim2demand}\vspace{-5mm}
\end{figure}
\fi

\section{Conclusion}\enlargethispage{3mm}

This work provides two ILP-models for network configuration applicable on quantum annealing computers. The work is based on the idea that NP-hard network optimization is more efficiently obtainable by adiabatic quantum computing compared to classical computing. Considering a wide-area network that has to face the challenge of bursty network traffic, a significant traffic loss reduction is obtainable if a resource reoccupation algorithm is applied every \SI{100}{ms}. This operation mode leads toward a zero-margin network as network resources are more efficiently used. Quantum computing might be a key enabler to achieve the proposed network operation in real-sized networks. 

The reduction of buffer lengths in networks as proposed in this work has a positive effect on the latency of data flows. This is especially the case for some use cases with latency requirements between 300ms to seconds. However, as wide-area networks span large distances up to thousands of kilometer, propagation delays of up to tens of milliseconds are expectable. Thus, for use-cases with strictly low latency requirements, e.g. below 20 ms, cloud computing solutions with server in the metro or regional area are preferable. 
Anyhow, as a higher meshing of metro networks is assumable within the future to realize high performance cloud computing networks, our approach may be also applicable there.\enlargethispage{3mm}

\end{document}